\documentclass[10pt,twocolumn,letterpaper]{article}

\usepackage{cvpr}
\usepackage{times}
\usepackage{epsfig}
\usepackage{graphicx}
\usepackage{amsmath}
\usepackage{amssymb}

\usepackage[breaklinks=true,bookmarks=false]{hyperref}

\cvprfinalcopy 

\def\cvprPaperID{****} 

\begin{document}

\title{Music Generation Using Deep Learning}

\author{Vasanth Kalingeri\\
University of Massachusetts Amherst\\
Amherst, Massachusetts\\
{\tt\small vkalingeri@cs.umass.edu}
\and
Srikanth Grandhe\\
University of Massachusetts Amherst\\
Amherst, Massachusetts\\
{\tt\small sgrandhe@cs.umass.edu}
}

\maketitle

\begin{abstract}
   The use of deep learning to solve the problems in literary arts has been a recent trend that gained a lot of attention and automated generation of music has been an active area. This project deals with the generation of music using raw audio files in the frequency domain relying on various LSTM architectures. Fully connected and convolutional layers are used along with LSTM's to capture rich features in the frequency domain and increase the quality of music generated. The work is focused on unconstrained music generation and uses no information about musical structure(notes or chords) to aid learning.The music generated from various architectures are compared using blind fold tests. Using the raw audio to train models is the direction to tapping the enormous amount of mp3 files that exist over the internet without requiring the manual effort to make structured MIDI files. Moreover, not all audio files can be represented with MIDI files making the study of these models an interesting prospect to the future of such models. 
\end{abstract}

\section{Introduction}

Music composition is an art, even the task of playing composed music takes considerable effort by humans. Given this level of complexity and abstractness, designing an algorithm to perform both the tasks at once is not obvious and would be a fruitless effort. It is thus easier to model this as a learning problem where composed music is used as training data to extract useful musicial patterns.

Our goal is to generate music that is pleasant to hear but not necessarily one that resembles how humans play music. We are hoping for the learning algorithm to find spaces where music sounds pleasant without enforcing any restrictions on whether it adheres to guidelines of musical theory. For this reason, we do not use any features such as notes, notation or chords to aid the learning or generation process, instead, we directly deal with the end result which are audio waveforms.

\begin{figure}[h]
\begin{center}
   \includegraphics[width=0.8\linewidth]{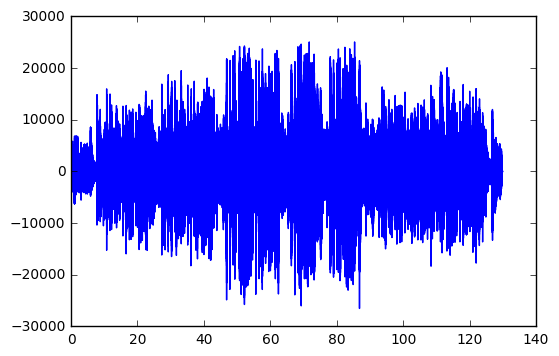}
\end{center}
   \caption{A simple visualization of sample audio waveform.}
\label{fig:long}
\label{fig:onecol}
\end{figure}

Audio waveforms are one dimensional signals, they vary with time such that an audio fragment at a particular timestep has a smooth transition from audio fragments from previous timesteps. Fig.1 provides a simple visualization of the raw audio waveform. An obvious choice of architecture to model a time varying function would be a recurrent neural network because of its ability to share parameters across time. Specifically, we will be using Long Short Term Memory(LSTM) networks to model the signals.

\section{Related work}

There has been a lot of work where musical features such as notes, chords and notations have been used to generate music using LSTMs[3,4,5]. These works show promising results and demonstrate that LSTMs have the ability to capture enough long-range information required for music generation. A typical architecture in these approaches involves structured input of a music notation from MIDI files that are encoded as vectors and fed into an LSTM at each timestep. The LSTM then predicts the encoding of the next timestep and this continues. The error is computed as negative log loss of the predicted value and the true value.

These approaches eliminate any possibility of noise as during generation, the predicted notes are mapped to their respective limited audio vocabulary in a one to one fashion. Because training is done on very low dimensional vectors, a higher cell size can be afforded by the LSTM there by increasing the time-range that they can learn. The choice of log loss also allows easier training of the architectures. For these reasons, these approaches are able to generate fairly pleasant audio. However, they are restricted to the kind of audio they generate because the outputs are restricted to the low dimensionality of input vector that the model operates on. These restrictions are completely removed when the networks deal with raw-audio information.    

There has also been work that directly models the raw audio samples, the best results are obtained from the paper[2] which uses a very deep dilated convolutional network to generate samples one at a time sampled at 16KHz. By increasing the amount of dilation at each depth they were able to capture long range dependencies to obtain a good representation of the audio. It is easy to imagine the kind of depth they had to attain in order capture enough time information that could encode the song well. Despite the large depth, training the network was relatively easy because they treated the problem as a classification problem where they classified the generated audio sample into one of 255 values. This allowed them to use the negative log loss instead of the mean squared loss. This reduced the likelihood of overfitting to outliers and thus decreased convergence time. Although this method works, the depth makes it extremely demanding in terms of computation, it takes about a minute to generate one second of audio on Google's GPU clusters forcing us to consider a faster alternative.

Recent work by Nayebi et. al[1] have also worked on audio samples, but instead of trying to learn and generate from raw audio samples, they work on the frequency domain of the audio. This approach is much faster because it allows the network to train and predict a group of samples that make up the frequency domain rather than one sample. Since the frequency domain can still represent all audible audio signals, there are no restrictions on the kind of  music it can generate. They use a single LSTM architecture, where the samples in the Fourier domain are fed as input at each timestep. The LSTM generates the output which is the Fourier representation of the signal of the next timestep. The mean squared difference between the predicted output and the true frequency representation is used as the cost function to train the network. Because of good results and lower computation demands, we implemented this method as the baseline for all comparisons.

We found several drawbacks in the method, although there was a good tune present in the audio samples, it was covered with a lot of disturbance, the network failed to capture long range dependencies and sounded pleasant only in some regions lacking a coherent structure. It is already known that an LSTM can be used to capture long-range dependencies as seen in [3, 4, 5], but the particularly large feature size and low cell state size imposed due to hardware restrictions prevent the network from doing so. We thus make changes in the architecture to mitigate these issues and generate more coherent music. 

\section{Approach}
All of the approaches deal with the frequency representation of the audio samples. The audio used is sampled at 16KHz and we only consider a single channel of audio for the sake of simplicity. Consider the raw audio samples to be represented as $<a_{0}, a_{1}, a_{2}, .... a_{n}>$, the Fourier transform is obtained by considering $n$ of these samples at a time to obtain a $n$ dimensional vector. Each of the values in the $n$ dimensional frequency vector consist of real and imaginary components, to simplify calculations, the vector is unrolled as $2n$ dimensional vector (D) where the imaginary values are appended after the $n$ real values. The LSTM used in all the approaches are the vanilla LSTMs from [7].

The rest of the details are implementation specific, and are discussed in the following subsections.
\subsection{Baseline implementation}
We established the work done by [1] as the baseline to compare all the other models against. This paper was implemented and the network architecture for the same can be seen in figure 2.

\begin{figure}[h]
\begin{center}
   \includegraphics[width=0.8\linewidth]{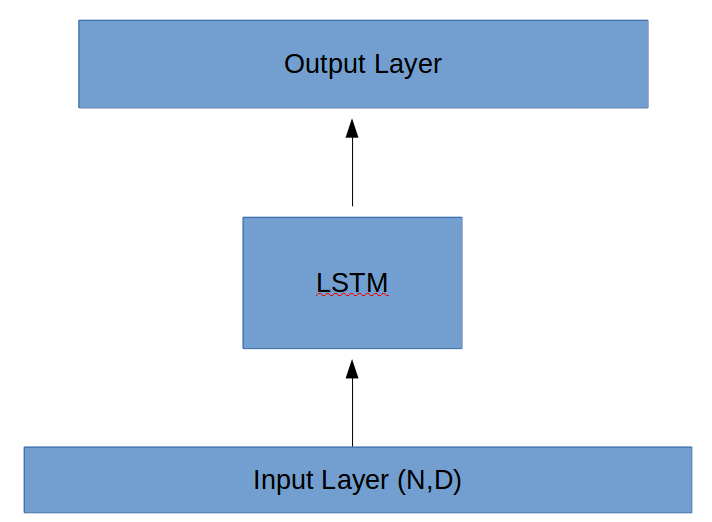}
\end{center}
   \caption{Base implementation.}
\label{fig:long}
\label{fig:onecol}
\end{figure}

The base model is fed a D dimensional input vector from training set to the LSTM which then generates the frequency vector of D dimensions as output. Consider the frequency representations to be represented by X$_{t}$ to denote the frequency component at the t$^{th}$ timestep. 

Consider T to be the number of timesteps that the LSTM is rolled over(which again is a hyperparameter like the cell state) during training ie. samples X$_{t0}$ till X$_{t0 + T}$ are fed as a sample in a batch for the LSTM at each timestep i allowing the LSTM generate output $\bar{X_{ti + 1}}$. The Mean Squared Error generated between X$_{ti + 1}$ and the true sample X$_{ti + 1}$ is calculated and is backpropagated through the LSTM. This process is carried at every timstep in the reverse order starting from X$_{t0 + T}$ to X$_{t0}$. During prediction, an initial sequence of T timesteps is required as input to generate the output at T+1 timestep which forms the first timestep of the predicted audio. This value is appended to the input sequence and the last T steps of the sequence are considered again for the next step of prediction. More formally, starting with X$_{0}$ to X$_{t}$, the value $\bar{X_{t + 1}}$ is predicted, to obtain the next prediction value X$_{1}$ to X$_{t + 1}$ where X$_{t + 1}$ = $\bar{X_{t + 1}}$ and this process is repeated.

We observed that predictions made in this fashion resulted in better sounding outputs than using predictions at each timestep and feeding them as input to the LSTM. This is because in this method each prediction has the context from all the T timesteps. However, if we stuck with the latter method, then the first predicted value starts with no context, the second prediction only has the context of the previous value and so on. Predictions made with small context are more error prone and tend to be random, this randomness cascades through time resulting in a poor result. Therefore, a sample X$_{0}$ to X$_{t}$ is randomly chosen as the starting seed and predictions are made using them.

From experiments we found the optimal value of D (sample dimensionality) to be 8000 and the value of the hidden state to be 2048. The fact that good music was generated demonstrates that the 2048 vector not only holds enough information about how to operate on the current state but also contextual information regarding how transitions are to be made from different states. The cell state decides how the hidden state vector has to be populated to pass relevant contextual information. Thus, the lack of long-range dependency could be naively solved by increasing the size of the cell state, however, due to hardware limitations, a good increase became unfeasible.

The reduction from 8000 dimensional vector to a space less than 2048(since we can assume some of the dimensions are storing contextual information) hints that only a subset of these frequencies are essential for human perception[1] and that the cell state stores a linear combination of the frequencies that are required for good prediction. If the cells could instead store a more complex representation of the frequencies, it would require relatively less dimensions to hold this information. We can expect to increase this complexity without experiencing considerable loss in performance because we know that not all of the frequency components are essential. So if the input to the LSTM was a complex function of frequencies, we could essentially get away with storing them in a lower dimensional cell state, leaving more room to store contextual information. The basis of this hypothesis led us to our second approach which involved adding layers in between the LSTM and the input, with the hope to allow the cell state to hold more contextual information and thereby increase the long-range dependency.

\subsection{Feed complex representations as input to the LSTM}
We confirmed our hypothesis experimentally by trying two different architectures as shown in figure (3, 4). Due to the lack of any specific structure to bank on in the case of music, we could not test them as concretely as done by Andrej Karpathy et al[6]. 

\begin{figure}[h]
\begin{center}
   \includegraphics[width=0.8\linewidth]{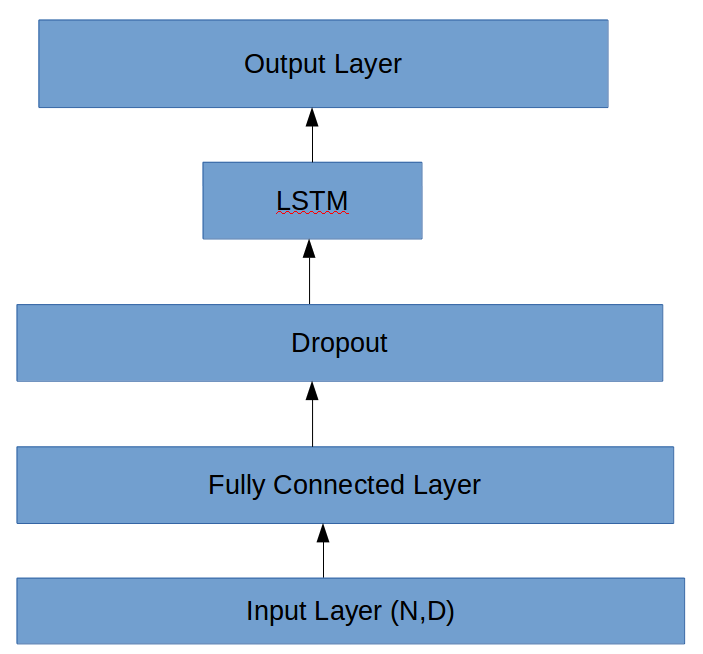}
\end{center}
   \caption{Fully connected architecture at a single timestep.}
\label{fig:long}
\label{fig:onecol}
\end{figure}

\begin{figure}[h]
\begin{center}
   \includegraphics[width=0.8\linewidth]{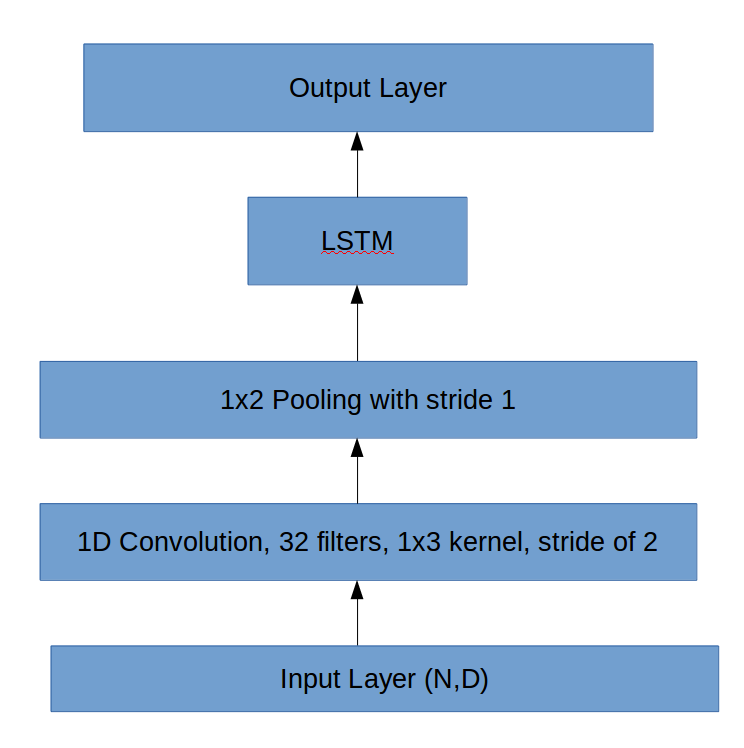}
\end{center}
   \caption{1D convolution architecture at a single timestep.}
\label{fig:long}
\label{fig:onecol}
\end{figure}

The architecture (fig 3.) involves passing the input at each timestep through a fully connected layer and feeding the output of this layer to the LSTM. The LSTM is then used to make predictions. Rectified Linear Units(RELUs) is used as the non-linearity in the fully connected layer. Experimentally we found that, setting the size of the fully connected layer equal to the size of the LSTM hidden state resulted in the best predictions. 

Looking at spectrographs of real music we can observe good local dependencies among various frequencies hinting that convolutional layers can be used instead of the fully connected layers giving us complex representations with fewer parameters. We added a 1D convolutional layer followed by a max pooling to the input layers(reference architecture with Conv 1D and max pooling), 12 filters of size 1x3 were used with a 2x2 max pooling. This operation was performed at every timestep and its output was fed into the LSTM. This decreased training times(reference plot on how it decreased) but the quality of music generated was not as good showing that the frequencies exhibit non-local dependencies among themselves which cannot be visually observed.

\begin{figure}[h]
\begin{center}
   \includegraphics[width=0.8\linewidth]{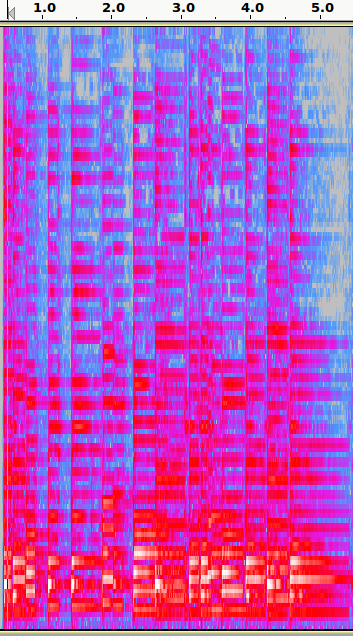}
\end{center}
   \caption{Real spectogram}
\label{fig:long}
\label{fig:onecol}
\end{figure}

\subsection{Using multiple LSTMs}
Increasing the complexity of the input was not enough to capture long-range information like that observed in research works that used MIDI files, so we wanted to increase the complexity of storing contextual information as well. We did this by implementing two different architectures that used multiple LSTMS.

\begin{figure}[h]
\begin{center}
   \includegraphics[width=0.8\linewidth]{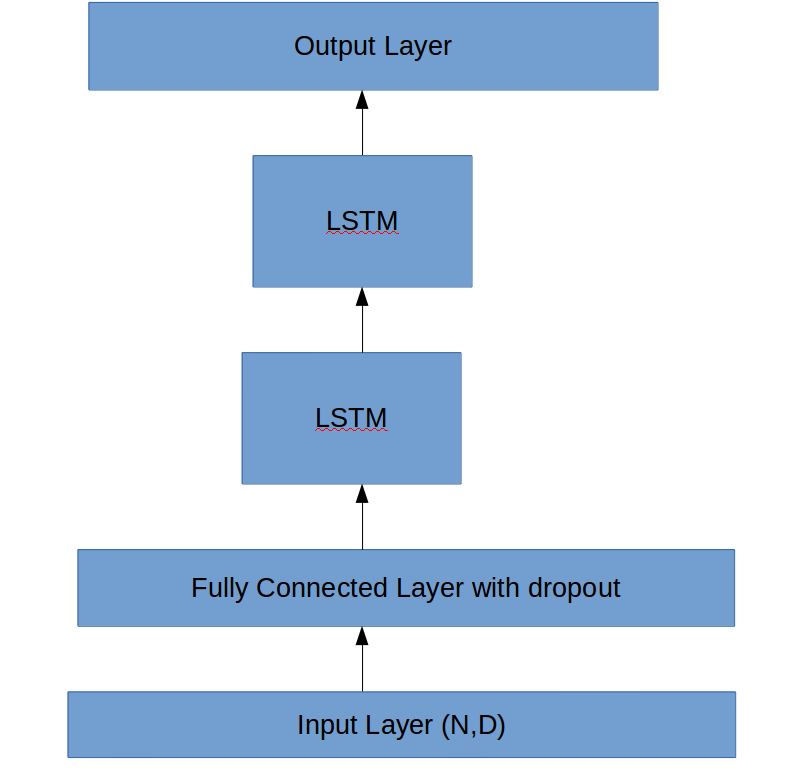}
\end{center}
   \caption{LSTMs stacked vertically (Multilayer LSTM).}
\label{fig:long}
\label{fig:onecol}
\end{figure}

The first approach involved stacking an LSTM in sequence to the first LSTM to incorporate more information in the hidden states at each timestep. The output of the first LSTM would be passed as input to the next LSTM, the second LSTM would output the prediction at each timestep. 

\begin{figure}[h]
\begin{center}
   \includegraphics[width=0.8\linewidth]{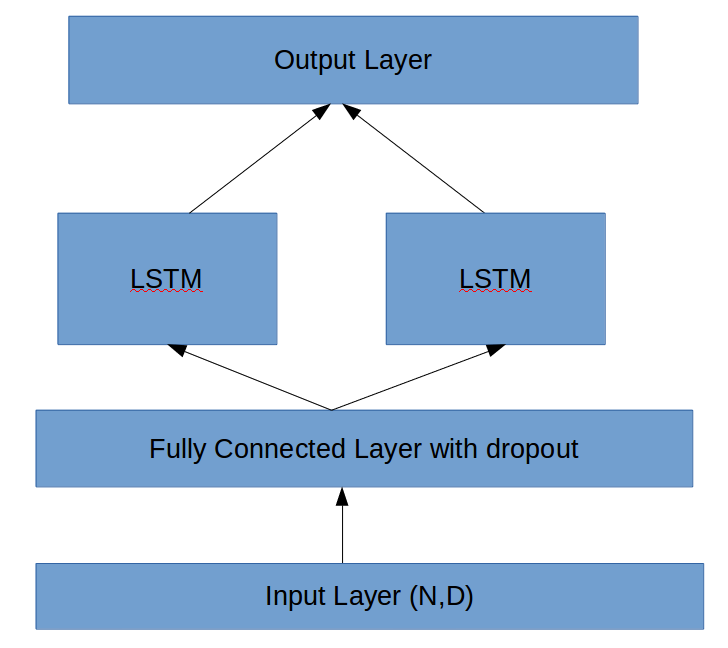}
\end{center}
   \caption{LSTMs in a bilinear alignment.}
\label{fig:long}
\label{fig:on ecol}
\end{figure}

In the second approach a bilinear model is constructed, as shown in figure(reference the figure) where two LSTM's read the input from the fully connected layer at each timestep and the merged output is fed to a fully connected layer. The merge layer just sums the results from both the layers and feeds it into a fully connected layer which then makes the predictions. The sum operation allows gradients to flow easily during backpropagation. Both the LSTMs are initialized randomly, so this way both the cell states are forced to learn different representations.

The intuition behind the bilinear model can be accounted due to the fact that since each LSTM learns only a fraction of the features in comparison to what a single LSTM models needs to learn, the architecture allows the LSTM's to learn faster in parallel and share the parameters among each other. We also tried an ensemble of stacked LSTM models with very low dimensional cell states (due to hardware constraints) without any convincing results.

\subsection{Adding convolutions to increase range}
We wanted to explore if we could further increase the range of context considered by the LSTM. We observe that when LSTMs are trained with MIDI files, each step of the LSTM receives a notation that translates to about 0.5 seconds of information or more.We presume that if the LSTM could operate on a larger time instance at once, it would be much easier to remember states since it can rely on a lot of context to come from the input itself. However, directly using more than 0.5 seconds of audio in the frequency domain results in 16000 dimensional vector that cannot be trained since it requires higher cell state size and more memory. We instead resorted to the architecture shown in figure 8.

\begin{figure}[h]
\begin{center}
   \includegraphics[width=0.8\linewidth]{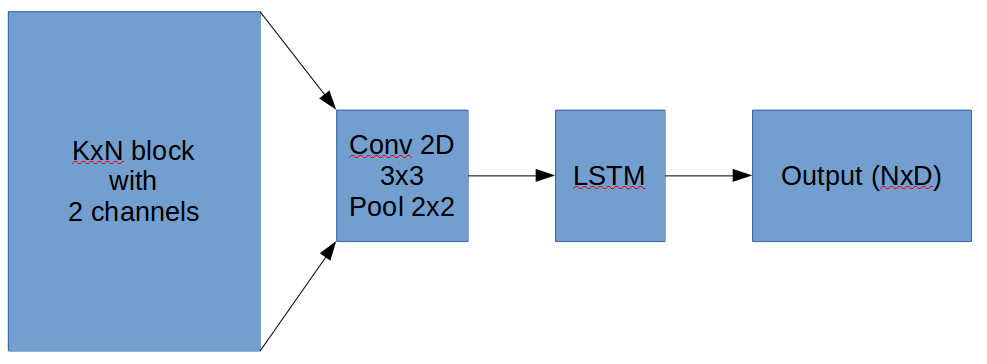}
\end{center}
   \caption{Feeding K blocks of N samples at single timestep. }
\label{fig:long}
\label{fig:onecol}
\end{figure}

In this method, the input representation is slightly modified, two channels are created of the frequency representation where the first channel holds the real values and the second channel holds the imaginary values, creating a two channeled N dimensional matrix. Instead of considering only one set of N samples at a time, we consider K groups of such two channeled N dimensional vectors per time step. The groups are stacked sequentially creating a KxN image with two channels where the LSTM generates the (k + 1)$^{th}$ sample as output at each timestep. The image(reference the image used) is very similar to the process of generating a spectograph. This is fed to a convolutional network, the output of which is fed to an LSTM that performs the prediction. The CNN here uses the spatial information of the these images and feeds in a compressed representation of that for the LSTM to analyze, this way we are able to feed in 0.75 seconds of info to an LSTM at a single timestep.

The music generated from this method is pleasant to hear, however, it was seen that the ones generated from bilinear LSTMs were better in most cases. 

\section{Experiments}
As part of dataset collection, piano tunes were scraped from the website $http : //orangefreesounds$ as mp3 files. These files were then converted to mono channel wav files sampled at 16KHz using ffmpeg. The dataset consists of 25 piano tunes of duration varying between 2 minutes to 3 minutes. 20 songs in the collected set are randomly picked for the training phase and remaining 5 songs are used as random starting seeds for prediction. In order to construct the dataset appropriate for the models, each of the songs are sampled at 0.25 seconds and converted to frequency domain using Fourier transform to obtain 4000 frequency components. Each of the frequency components consists of both real and the imaginary part which when unrolled give us an 8000 dimensional vector where the imaginary values are appended to the real values.

The 8000 dimensional vector forms the input to the model at each timestep. The model has been trained to utilize 40 timesteps worth of information for each iteration allowing 10 seconds worth of information to be processed by the model. Each of the audio file is segmented into 10 second intervals giving nearly 300 samples for training the model.

The models were implemented and executed using the Keras wrappers with Tensorflow framework as the backend to run the models on the Nvidia GTX960M GPU.

Each of the models was trained for 2000 epochs using the rmsprop optimizer and a learning rate of 0.0001. The Mean Square Error function used for designing the loss function for all the models. For the Fully connected layers in the models combination of L2 regularization and dropout was used. The L2 values were tested with 0.0001 along with dropout values of 0.2. For the LSTM models dropouts of 0.5 was used to provide regularization for the hidden state weight matrix and the input weight matrix. The plots below show the drop in training loss with respect to the epochs:

\begin{figure}[h]
\begin{center}
   \includegraphics[width=0.8\linewidth]{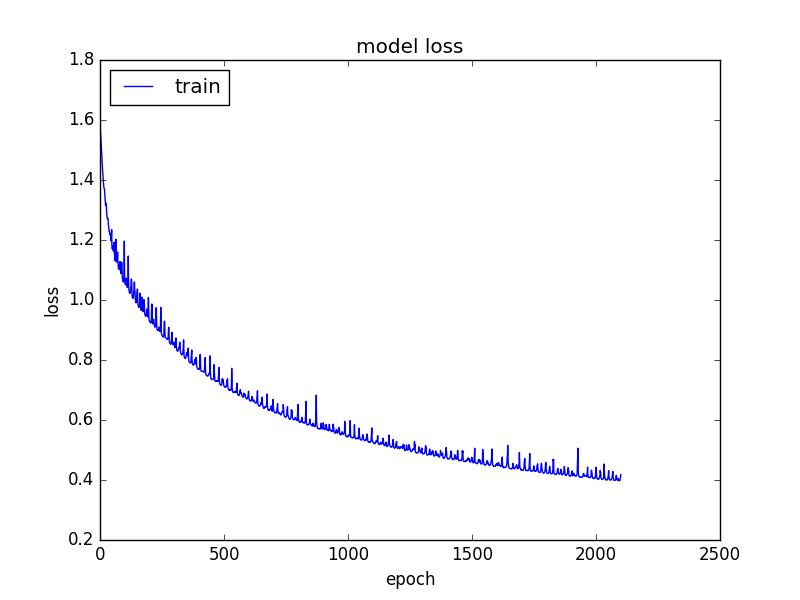}
\end{center}
   \caption{Loss plot of Bilinear LSTM}
\label{fig:long}
\label{fig:onecol}
\end{figure}

\begin{figure}[h]
\begin{center}
   \includegraphics[width=0.8\linewidth]{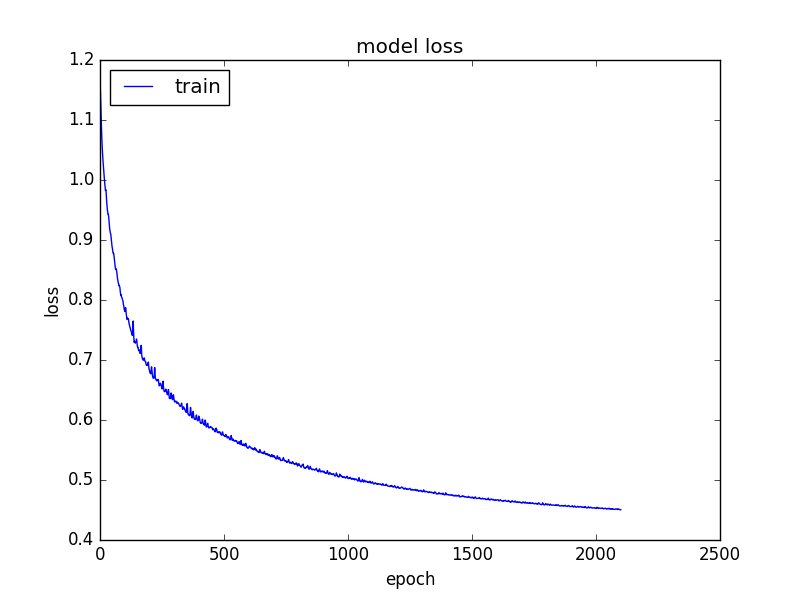}
\end{center}
   \caption{Loss plot of the base implementation}
\label{fig:long}
\label{fig:onecol}
\end{figure}

The loss plots of all the implementations look very similar and they play no role in deciding the quality of generated music, they were only used to aid in the selection of hyper-parameters of the network.

Each of the trained model is then subject to the prediction phase where the models are fed with random sequences from the test set to generate 10 second audio clip with the weights generated at different epochs. The samples were generated at 1200, 1500, 1800 and 2100 epochs to pick the best model possible. The generated clips are then pruned manually to create a cluster of the plausible audio from different models.

Blind testing approach was used for gathering the ratings on generated audio samples. A collection of 30 audio samples was collected from different models and presented to 10 testers to provide rating for each audio sample on a scale from 1 to 5. The aggregated results obtained can be found in Table 1.

\begin{table}
\begin{center}
\begin{tabular}{|l|c|}
\hline
Model & Average Rating \\
\hline\hline
Base Model & 2.4 \\
FC with LSTM & 3.2 \\
2D-Conv with LSTM & 3.4\\
1D-Conv with LSTM & 3.0 \\
Bilinear LSTM Model & 3.6 \\
Multilayer LSTM (stacked Model & 2.8 \\
\hline
\end{tabular}
\end{center}
\caption{Results of blind testing performed on 10 testers with 5 generated samples picked from each model.}
\end{table}

All the proposed models show to outperform the base model proposed in [1] with the Bilinear model providing the best output samples of all the models. The bilinear model allows the LSTM's to learn in parallel from scratch by achieving the performance of the other models such as the RNN with FC layer at a lower number of epochs. Since the outputs of the LSTM's in the bilinear model get summed up each of the LSTM's can converge faster alowing them to capture more features from the dataset.

The multilayer LSTM models, the single LSTM with fully connected layer and the single LSTM with convolution layers all outperform the base model. This could be credited to the fact that all these modesl have more parameters than the base allowing the models to learn better from the limited training data.

Though Piano tunes were used to train the model, the expected output from the model is not to generate Piano tunes in return. The expectation is to generate an audio that is pleasing to the ear which may or may not be intrumental tones. The model uses its ability of propagating information at different frequencies to artficially synthesize tunes which are more or less electronically generated using combination of multiple frequencies and varying amplitudes.     

\section{Conclusion}
The models studied showed improved performance in the quality of music over the base model and provides more insights into the architectures that would suit raw audio representations. The bilinear architecture and the LSTM with 2D convolutional layers produced the best audio of the tested models.  

In future, we would like to investigate the option of changing the loss function from the MSE to using an adversarial network and allow the network to model its own loss function based on the available data, one that can better represent how humans perceive music.

\section{References}

\begin{enumerate}
	\item Nayebi,  A.,  and  Vitelli,  M.   2015.   GRUV  :  Algorithmic
Music Generation using Recurrent Neural Networks
    \item Aaron van den Oord, Sander Dieleman, Heiga Zen, Karen Simonyan, Oriol Vinyals, Alex Graves, Nal Kalchbrenner, Andrew Senior, Koray Kavukcuoglu  WaveNet: A Generative Model for Raw Audio  arXiv:1609.03499v2, 2016
    \item Huang, A., and Wu, R,   Deep Learning for Music arXiv:1606.04930v1,2016
	\item Daniel Johnson. Composing music with recurrent neural networks.
$https : //goo.gl/YP9QyR$
	\item Google Magenta $https : //magenta.tensorflow.org/$
    \item Andrej Karpathy, Justin Johnson, and Li Fei-Fei. Visualizing and understanding recurrent networks.In ICLR 2016 Workshop, 2016
    \item Sepp Hochreiter, Jurgen Schmidhuber, Long short-term memory: $http://deeplearning.cs.cmu.edu/pdfs/Hochreiter97_lstm.pdf$
\end{enumerate}

\end{document}